\documentclass[ reprint,
superscriptaddress,
amsmath,amssymb,
aps,
nofootinbib]{revtex4-2}

\usepackage{graphicx}
\usepackage{dcolumn}
\usepackage{bm}

\usepackage{xcolor}
\usepackage{mathrsfs}
\usepackage{mathtools}

\begin{document}
\preprint{APS/123-QED}

\title{Moving Manifolds and General Relativity}

\author{David V. Svintradze}
\email[Correspondence email address: ]{dsvintradze@newvision.ge}

\affiliation{School of Medicine, New Vision University, Bokhua 11, 0159 Tbilisi, Georgia}

\affiliation{Niels Bohr Institute, University of Copenhagen, Blegdamsvej 17, 2100 Copenhagen, Denmark}

\date{\today}

\begin{abstract}
We revise general relativity (GR) from the perspective of calculus for moving surfaces (CMS). While GR is intrinsically constructed in pseudo-Riemannian geometry, a complete understanding of moving manifolds requires embedding in a higher dimension. It can only be defined by extrinsic Gaussian differential geometry and its extension to moving surfaces, known as CMS. Following the recent developments in CMS, we present a new derivation for the Einstein field equation and demonstrate the fundamental limitations of GR. Explicitly, we show that GR is an approximation of moving manifold equations and only stands for dominantly compressible space-time. While GR, with a cosmological constant, predicts an expanding universe, CMS shows fluctuation between inflation and collapse. We also show that the specific solution to GR with cosmological constant is constant mean curvature shapes. In the end, by presenting calculations for incompressible but deforming two-dimensional spheres, we indicate that material points moving with constant spherical velocities move like waves, strongly suggesting a resolution of the wave-corpuscular dualism problem. 
\end{abstract}

\maketitle

\textit{Introduction--} General relativity (GR) is the most successful theory of gravitation \cite{Will:2014aa, PhysRevLett.123.191101}. Indeed, experimental tests (to name a few) performed in the nonlinear and dynamical regimes produced a high degree of accuracy \cite{PhysRevD.80.122003, PhysRevD.100.104036}. Recent extensions of GR, among many others, are embedding the brane in higher dimensional space \cite{DVALI2000208} and modifications of Einstein-Hilbert (EH) action with quadratic terms \cite{ALEXANDER20091}. It is noteworthy that recent experiments constrain largely deviating from GR new theories  \cite{PhysRevLett.123.191101}, but there is an experimental indication of the necessity of extending GR fundamentally \cite{PhysRevLett.131.111001}. 

Here, we refrain from making any adjustments in the field and, adhering to mathematically rigorous notations, we label actions as
\begin{align}
S_\Omega&=\int_\Omega\mathscr{L}_\Omega d\Omega \label{space} \\
S_\mathbb{S}&=\int_\mathbb{S}\mathscr{L}_\mathbb{S}d\mathbb{S} \label{surface}
\end{align}
where $\Omega$ stands for the bounded by $\mathbb{S}$ hyper-surface space for embedded manifold so that 
\begin{equation}
d\Omega=\sqrt{\eta}d^{n+1}X=\sqrt{\eta}dX^1 \cdots   dX^{n+1} \nonumber
\end{equation} 
$\eta=\pm\det(\eta_{ij}) $ is the determinant of the space metric tensor with the sign dependently whether space is Riemannian or pseudo-Riemannian. The hyper-surface area element is defined as
\begin{equation} 
d\mathbb{S}=\sqrt{g}d^nx \nonumber 
\end{equation} 
sign of $g$ depends on whether the manifold is Riemannian or pseudo-Riemannian. Later, we omit "hyper" and use surface or manifold for all dimensional hyper-surfaces. We designate the space metric tensor as $\eta$ and the surface metric as $g$.
As far as the surface and enclosed space are evolving and continuously changing, we evaluate (\ref{space}, \ref{surface}) actions according to the extension of differential geometry called calculus for moving surfaces (CMS). We do not modify fields $\mathscr{L}_\Omega,\mathscr{L}_\mathbb{S}$ and, as a primary example, work with the EH action:
\begin{equation}
\mathscr{L}_\mathbb{S}=\frac{1}{2k}R+\sigma \label{EH action}
\end{equation} 
where $k=8\pi Gc^{-4}$ Einstein gravitational constant, $G$ is the gravitational constant, $c$ is the speed of light, $R$ is the Ricci scalar, and $\sigma$ is some surface energy density. Quadratic or any other modifications can be implemented at any stage due to the generality of our formalism. Because the manifold is moving (changing by some extra parameter), one needs an extrinsic approach to encounter the motion fully, embedding it in a higher dimension. Therefore, both (\ref{space}, \ref{surface}) space and surface actions stay relevant, though here we only concentrate on the surface action (\ref{surface}) to demonstrate alternative ways of taking variations and generalizing GR. We implement embedding in higher dimension space-time to consider the manifold's shape motion properly. Therefore, the presented formalism applies to all $\mathscr{L}_\Omega,\mathscr{L}_\mathbb{S}$ space and surface energy densities. Because of independence from causality, the manifold is assumed to be moving, 
the actions are 
\begin{align}
S_\Omega&=\int_\mathbb{S}\frac{\rho V^2}{2}d\mathbb{S}-\int_\Omega\mathscr{L}_\Omega d\Omega \label{space action} \\
S_\mathbb{S}&=\int_\mathbb{S}\frac{\rho V^2}{2}d\mathbb{S}-\int_\mathbb{S}\mathscr{L}_\mathbb{S}d\mathbb{S} \label{surface action}
\end{align}
where $\rho$ is the surface mass density (initially can be zero), $V$ is the surface velocity, and $\mathbb{S}\in\Omega$, therefore the most generic action is (\ref{space action}). The action is (\ref{space action}) if the surface is embedded in higher dimension space and the $\mathscr{L}_\Omega$ is the space energy density\footnote{energy per unit volume, in three dimensions is pressure.}. While the action is (\ref{surface action}) if the surface dynamics can be described by the surface energy density $\mathscr{L}_\mathbb{S}\in\mathscr{L}_{\Omega}$\footnote{energy per unit area, in three dimensions is tension.} only. Following the definitions (\ref{space action}, \ref{surface action}), we consider the variation of kinetic terms to be a variation of Lagrangian terms and all the modeling of $\mathscr{L}_\mathbb{S}, \mathscr{L}_\Omega$ serves to capture kinetic energies of the surface motion. Since here we deal with mostly EH action where there is no direct higher dimensional embedding, we mostly work with $S_\mathbb{S}$ (\ref{surface action}) and invoke space terms when needed explicitly. In this formalism, we show that the surface neither expands nor collapses for dominantly compressible space-time but oscillates non-harmonically. Therefore, all measurements done in the expansion regime will stay consistent with GR in some approximations. 

Our motivation is simple. The Riemannian geometry is intrinsically defined as GR.  More implicitly, one can introduce a metric tensor without invoking tangent space and construct all differential geometry based on partial derivatives of it (see below). For instance, Christoffel's symbols become the combination of metric tensor partial derivatives and are renamed by Levi-Civita connections, while the combination of partial derivatives of the connections gives Riemann tensor. Metric and Riemann tensors are only intrinsic fundamental tensors, and all other intrinsic tensors are given by the combination of the two. While Riemannian geometry is the perfect language to describe the not-moving manifold (static geometry), it cannot describe shape motion, which is the manifold that evolves with some extra parameters. Throughout the paper, we refer to the additional parameter as time, bearing in mind that the extra parametric time might have nothing to do with the real or proper time we use in physics. To distinguish the two, we designate the parametric time as $t$ and the real/proper time as $\tau$. In some approximations, parametric and real times can be equated $t=\tau$ and, therefore, consequent equations simplified, though it is not our goal in this paper. In GR, $\tau$ proper time is considered a pseudo coordinate, and consequently, the space-time manifold becomes a four-dimensional hyper-surface. The static manifold is considered to be moving because of the presence of $\tau$ proper time in the metric tensor. The causality of this assumption is that a-priory statically formulated Riemannian geometry is used to describe equations of motions for moving manifolds. In other words, it is "conjectured/hypothesized"  that static geometry can still describe moving manifolds.  We challenge this idea here and, quite contrary, show that to understand all possible deformations of moving manifolds fully, one needs an extension of differential geometry called the calculus of moving surfaces (CMS) \cite{Grinfeld2013}. Using CMS, we generalize GR and show that space-time fluctuates with inflation and collapses in the dominantly compressible universe, while the surface mass density non-harmonically, with growing amplitudes, oscillates locally. In a predominantly incompressible universe, it is possible to form various shapes with different patterns and varying pressure and temperature distributions. Calculations in CMS formalism show that material points moving with constant spherical velocities behave like waves due to surface dynamics in two-dimensional incompressible spheres. While calculations are performed for simple two-dimensional surface cases, the generality of the method allows it to be readily extended to any pseudo-Riemannian geometry of any dimensions.    

\textit{Calculus for Moving Surfaces (CMS) --} Before proceeding with analyses, we briefly define the extension of differential geometry called the calculus for moving surfaces. The extension was initiated by Hadamard \cite{Hadamard} and developed by a generation of mathematicians \cite{Thomas, MGrinfeld, Grinfeld2013}. 

Using the CMS in its current form, P. Grinfeld proposed equations of motions for massive thin fluid films \cite{Grinfeld2013, Grinfeld2010, PhysRevLett.105.137802, Grinfeld2012}. We have generalized P. Grinfeld equations for any manifolds and derived surface dynamics equations \cite{Svintradze2017, Svintradze2018}, solved shape-dynamics problems for water drops \cite{Svintradze2019}, and extended Young-Laplace, Kelvin, and Gibbs-Thomson equations to arbitrarily curved surfaces \cite{Svintradze2020, Svintradze2023}, see also meeting abstracts that preceded the papers \cite{Svintradze:2024aa, Svintradze2023a, Svintradze2022a, Svintradze2021a, Svintradze2020a, Svintradze2017a, Svintradze2016a, Svintradze2015a, Svintradze2014a, Svintradze2013a, Svintradze2011a, Svintradze2009a}.

\textit{Gaussian Differential Geometry --} Here, we revive the concept of extrinsic differential geometry, referred to as Gaussian geometry. The approach developed by Gauss, in contrast to Riemannian geometry, is fundamentally extrinsic. We shall soon see that while Riemannian geometry is a powerful tool for describing static surfaces, it is not enough to encounter moving shapes entirely. Embedding in higher dimensions is necessary to better understand moving manifolds. Therefore, instead of describing Riemannian geometry, we start from the Gaussian one, bearing in mind that the formalism is also sufficient for higher than two-dimensional surfaces. In mathematics, higher dimensional surfaces are called hyper-surfaces. To minimize terminology, we refer to any dimensional manifold as a surface and explicitly indicate whether the referred surface is two-dimensional. 

Let $S^\alpha\in\mathbb{S}^n, \alpha=0,1,2,...,n,X^i\in\mathbb{M}^{n+1}, i=0,1,2...,n+1$ be the generic coordinates of the point on the hyper-surface $\mathbb{S}$ (referred as surface) embedded on one higher dimensional Minkowski space $\mathbb{M}$ so that $\bm{R}$ is a position vector $\bm{R}\in\mathbb{M}^{(n+1)}, \bm{R}=\bm{R}(X^i, t)=\bm{R}(S^\alpha, t)$. In contrast to our previous publications, greek indexes here stand for surface and Latin ones for ambient Minkowski space with the space-like signature $(-1,1,...,1)$, bold letters indicate vectors. Partial derivatives of the position vector define covariant base vectors of ambient space and embedded surface, while the scalar product of base vectors are metric tensors
\begin{equation}
\begin{matrix*}[l] 
\bm{X}_i=\partial_i\bm{R} & \bm{S}_\alpha=\partial_\alpha\bm{R} \\
\eta_{ij}=\bm{X}_i\cdot \bm{X}_j & g_{\alpha\beta}=\bm{S}_\alpha\cdot \bm{S}_\beta
\end{matrix*} \label{definition}
\end{equation}
where $\partial_i=\frac{\partial}{\partial X^i}, \partial_\alpha=\frac{\partial}{\partial S^\alpha}$. Note here that since in Minkowski space $\eta_{00}=-1$ according to the definition (\ref{definition}) $\bm{X}_0^2=-1$ and therefore space-time is defined in complex space, that makes the embedded surface pseudo-Riemannian manifold, therefore space and surface metrics are defined in complex space so that $g^2=gg^*=-g, \eta^2=\eta\eta^*=-\eta$. Inverse to space $\eta_{ab}$ and surface $g_{\alpha\beta}$ tensors are contravariant metric tensors $\eta_{ab}\eta^{bc}=\delta_a^c, g_{\alpha\beta}g^{\beta\gamma}=\delta_\alpha^\gamma$, where $\delta_\alpha^\gamma$ is the Kronecker delta matrix. The surface normal $\bm{N}$ is defined as the normal vector to surface bases so that $\bm{N}\cdot \bm{S}_\alpha=0, \bm{N}^2=1$.     

The definition of base vectors allows the introduction of Christoffel symbols for ambient and embedded spaces. For ambient Minkowski flat space, the symbols $\Gamma_{ij}^k=\bm{X}^k\cdot\partial_i\bm{X}_j=0$ are vanishing, while for embedded surface $\Gamma_{\alpha\beta}^\gamma=\bm{S}^\gamma\cdot\partial_\alpha\bm{S}_\beta\neq 0$ in general. In Riemannian geometry, the Christoffel symbols can be presented as a linear combination of partial derivatives of metric tensors. Therefore, the symbols are intrinsic
\begin{align}
\Gamma_{ij}^k&=\frac{1}{2}\eta^{kn}(\partial_j\eta_{in}+\partial_i\eta_{nj}-\partial_n\eta_{ij}) \label{ambient CR} \\
\Gamma_{\alpha\beta}^\gamma&=\frac{1}{2}g^{\gamma\mu}(\partial_\beta g_{\alpha\mu}+\partial_\alpha g_{\mu\beta}-\partial_\mu g_{\alpha\beta}) \label{surface C}
\end{align}
While all symbols are unchanged in differential geometry for moving manifolds, they are not the only ones and are enhanced by extrinsic parameters. Christoffel symbols allow the definition of invariant to reference frame covariant derivatives
\begin{align}
\nabla_k T_i^j=&\partial_k T_i^j - \Gamma^m_{ki} T_m^j + \Gamma_{mk}^j T_i^m \label {ambient C} \\
\nabla_\gamma T_{\alpha i}^{\beta j}=&\partial_\gamma T_{\alpha i}^{\beta j} +\eta_\gamma^k\Gamma_{km}^j T_{\alpha i}^{\beta m}-\eta_\gamma^k\Gamma_{ki}^m T_{\alpha m}^{\beta j} \nonumber \\
&+\Gamma^\beta_{\gamma\nu} T_{\alpha i}^{\nu j}-\Gamma^\nu_{\gamma\alpha} T_{\nu i}^{\beta j} \label{mixed C}
\end{align}
where $\eta_\gamma^k$ is the shift tensor, reciprocally shifting space/surface bases so that
\begin{equation}
\bm{S}_\alpha=\eta_\alpha^i\bm{X}_i, g_{\alpha\beta}=\bm{S}_\alpha\cdot\bm{S}_\beta=\eta_\alpha^i\bm{X}_i\eta_\beta^j\bm{X}_j=\eta_\alpha^i\eta_\beta^j\eta_{ij} \label{shift tensors}
\end{equation}
Note here that in flat Minkowski space-time, Christoffel symbols with Latin indexes vanish, therefore ${\nabla_k T_i^j=\partial_k T_i^j}$ and equations (\ref{ambient C}, \ref{mixed C}) simplify accordingly.   

Definitions (\ref{ambient C}--\ref{shift tensors}) provide metrilinic property of metric tensors $\nabla_k\eta_{ij}, \nabla_\gamma g_{\alpha\beta}=0$, consequently the surface base vectors are orthogonal to their covariant derivatives ${\bm{S}_\alpha\cdot\nabla_\gamma\bm{S}_\beta=0}$, therefore
\begin{equation}
\nabla_\alpha\bm{S}_\beta=\bm{N}B_{\alpha\beta} \label{curvature tensor}
\end{equation} 
$\bm{N}$ is the surface normal and $B_{\alpha\beta}$ is the curvature tensor and according to definition (\ref{curvature tensor}) the curvature tensor is symmetric and extrinsic parameter. Trace of the mixed curvature tensor $B_\alpha^\alpha$ is the mean curvature, while the determinant is the Gaussian curvature $K=\lvert B^\cdot_\cdot \rvert$.  

Note that, according to covariant derivatives and definitions of Christoffel symbols (\ref{ambient CR}--\ref{mixed C}), covariant derivatives are intrinsic, and therefore Riemann tensor $R^\delta_{\alpha\beta\gamma}$ defined as 
\begin{align}
&(\nabla_\alpha\nabla_\beta-\nabla_\beta\nabla_\alpha)T_\gamma=R^\lambda_{\alpha\beta\gamma}T_\lambda \label {Riemann} \\
&R^\rho_{\sigma\mu\nu}=\partial_\mu\Gamma^\rho_{\nu\sigma}-\partial_\nu\Gamma^\rho_{\mu\sigma}+\Gamma^\rho_{\mu\alpha}\Gamma^\alpha_{\nu\sigma}-\Gamma^\rho_{\nu\alpha}\Gamma^\alpha_{\mu\sigma} \nonumber
\end{align}
is also intrinsic. The Trace of the Riemann tensor is the Ricci tensor, and a trace of the mixed Ricci tensor is the Ricci scalar. 
\begin{align}
R^\delta_{\alpha\delta\gamma}&=R_{\alpha\beta} \\
R_\alpha^\alpha&=R
\end{align}
Extrinsic curvature tensor is related to intrinsic Riemann tensor according to \textit{Gauss's Theorema Egregium} 
\begin{equation}
R_{\gamma\delta\alpha\beta}=B_{\alpha\gamma}B_{\beta\delta}-B_{\beta\gamma}B_{\alpha\delta} \label{Riemann T}
\end{equation}
Therefore, intrinsic $\nabla, g, R$ tensors can describe all static surfaces. Note here that from the Gauss theorem (\ref{Riemann}), Ricci scalar follows
\begin{equation}
R=R_\alpha^\alpha=(B_\alpha^\alpha)^2-B_{\alpha\beta}B^{\alpha\beta} \label{Ricci}
\end{equation}
The contraction of the curvature tensor is negative twice Gaussian curvature for two-dimensional surfaces. 
\begin{equation}
2K=-B_{\alpha\beta}B^{\alpha\beta} \label{K definition}
\end{equation}
Therefore, the EH action resembles the Helfrich Hamiltonian for two-dimensional surfaces, which has been intensively studied in biomembrane applications. Note that (\ref{K definition}) can be readily generalized for higher dimensional manifolds as definition. 

Introducing the velocity field for moving manifolds clarifies the insufficiency of describing surface dynamics by only intrinsic tensors (see below).     

\textit{Velocity Field and CMS--} All definitions from Gaussian geometry (\ref{definition} -- \ref{curvature tensor}) are also applicable to moving manifolds. However, extensions are required to parameterize the motion because a surface changes. We refer to the parameter as time $t$ and assign $\tau$ to the proper time to distinguish the two. Parametric time might have nothing to do with real/proper time, but it is necessary to describe manifold motion. The question to be asked is whether or not the parametric time could be analyzed as relativistic real-time. That question we leave open here. 

The introduction of parametric time immediately necessitates defining time derivatives to preserve covariance. For that, we first introduce surface velocities: $C$ the surface normal velocity, and $V^\alpha$ tangent ones as components of ambient velocity 
\begin{equation}
\bm{V}=\partial_t \bm{R}=C\bm{N}+V^\alpha\bm{S}_\alpha \label{surface velocity}
\end{equation}
Note here that $\bm{V}, C, V_\alpha$ are only defined in ambient flat space and, therefore, are tensors only in the space. Note that the interface velocity normal component of the surface velocity is invariant 
\begin{equation}
C=\bm{N}\cdot\bm{V}=N^i\bm{X}_iV_j\bm{X}^j=N^iV_i \nonumber
\end{equation}
Therefore, if the surface is assumed to be moving initially, it will remain in all reference frames. The invariance of interface velocity straightforwardly indicates that, in general, for moving manifolds, a reference frame cannot be found where $\bm{V}$ vanishes. 

Introduction of the velocity field (\ref{surface velocity}), without further elaboration \cite{Svintradze2017, Grinfeld2013}, opens the path toward defining the covariant time derivative
\begin{align}
\dot\nabla T_\alpha^\beta&=\partial_tT_\alpha^\beta-V^\gamma\nabla_\gamma T_\alpha^\beta+\dot\Gamma_\gamma^\beta T^\gamma_\alpha-\dot\Gamma^\gamma_\alpha T_\gamma^\beta \label{covariant t} \\
\dot\Gamma_\alpha^\beta&=\nabla_\alpha V^\beta-CB_\alpha^\beta \label{Grinfeld symbol}
\end{align}
where (\ref{Grinfeld symbol}) $\dot\Gamma_\alpha^\beta$ is the Christoffel symbols analog for invariant time derivatives first introduced by P. Grinfeld \cite{Grinfeld2013}. $\dot\nabla$ is invariant and has all properties as covariant derivatives. From definitions (\ref{ambient C}, \ref{mixed C}, \ref{covariant t}, \ref{Grinfeld symbol}) straightforwardly follows that 
\begin{equation}
\nabla_\gamma g_{\alpha\beta},\dot\nabla g_{\alpha\beta}=0 \label{m property}
\end{equation} 
(\ref{m property}) is known as metrilinic property.
Taking into account (\ref{covariant t}--\ref{m property}) we have
\begin{align}
&\dot\nabla g_{\alpha\beta}=\partial_t g_{\alpha\beta}-V^\gamma\nabla_\gamma g_{\alpha\beta}-\dot\Gamma^\gamma_\alpha g_{\gamma\beta}-\dot\Gamma^\gamma_\beta g_{\gamma\beta} \nonumber \\
&=\partial_t g_{\alpha\beta}-(\nabla^\gamma V_\alpha-CB^\gamma_\alpha)g_{\gamma\beta}
-(\nabla^\gamma V_\beta-CB^\gamma_\beta)g_{\alpha\gamma} \nonumber \\
&=\partial_t g_{\alpha\beta}-(\nabla_\beta V_\alpha-CB_{\alpha\beta})-(\nabla_\alpha V_\beta-CB_{\beta\alpha}) \nonumber \\
&=\partial_t g_{\alpha\beta}-\nabla_\alpha V_\beta-\nabla_\beta V_\alpha+2CB^{\alpha\beta}=0, \Rightarrow \nonumber \\
&\partial_t g_{\alpha\beta}=\nabla_\alpha V_\beta+\nabla_\beta V_\alpha-2CB_{\alpha\beta} \label {metric derivative} 
\end{align}
For the contravariant metric tensor, (\ref{metric derivative}) has reversed signs: a negative sign applies for tangent velocities and a positive sign for normal velocity.

Even though the invariant time derivative (\ref{covariant t}) and covariant space derivatives (\ref{ambient C}, \ref{mixed C}) possess the same properties, they are fundamentally different. The first is extrinsically defined because it includes ambient surface velocities and the curvature tensor, whereas the latter is intrinsic. The \textit{Gauss's Theorema Egregium} (\ref{Riemann T}) essentially explains that static geometry can be intrinsically constructed. In contrast, equations (\ref{surface velocity}--\ref{Grinfeld symbol}) reveal that moving manifolds are essentially extrinsic phenomena unless a direct reasonable connection is made between extrinsically defined velocities and the intrinsically defined tensors. Additionally, it is worth noting that, in addition to (\ref{metric derivative}), the definitions (\ref{covariant t}, \ref{Grinfeld symbol}) lead to beautiful theorems \cite{Grinfeld2013}:
\begin{align}
\dot\nabla B_\alpha^\beta= \nabla_\alpha\nabla^\beta C+CB_\alpha^\gamma B_\gamma^\beta \label{covariant b} \\
\dot\nabla B_\alpha^\alpha= \nabla_\alpha\nabla^\alpha C+CB_\alpha^\beta B_\beta^\alpha \label{covariant m} 
\end{align}
(\ref{covariant m}) is the special case of (\ref{covariant b}). (\ref{covariant b}, \ref{covariant m}) theorems indicate that invariant time derivatives of the curvature tensor and the mean curvature are strongly dominated by the surface normal velocity $C$ and, therefore, cannot be intrinsically defined.
 
Definitions (\ref{ambient C}--\ref{curvature tensor}, \ref{surface velocity}, \ref{covariant t}) lead to integration theorems 
\begin {align}
\partial_t\int_\Omega fd\Omega&=\int_\Omega\partial_t f d\Omega+\int_\mathbb{S} fCd\mathbb{S} \label{space integral} \\
\partial_t\int_\mathbb{S} fd\mathbb{S}&=\int_\mathbb{S} \partial_t f d\mathbb{S}-\int_\mathbb{S}  fCB_\alpha^\alpha d\mathbb{S} \label{surface integral}
\end{align}
Equations (\ref{definition}--\ref{Grinfeld symbol}) along with integration theorems (\ref{space integral}, \ref{surface integral}) form the foundation of CMS and extend understanding of GR.
Proof of (\ref{space integral}) is known in mathematics. We provide a short proof of the (\ref{surface integral}) theorem here for brevity. 

Let $F\in \mathbb{S}$ be a smooth scalar function on the closed smooth manifold $\mathbb{S}$. Then, across any closed stationary contour $\gamma$, the contour integral vanishes because the contour normal $n_\alpha$ lying in the tangent plane always stays perpendicular to the surface tangent velocity $\bm{V}_T$, so that 
\begin{equation}
v=\bm{n}\cdot\bm{V}_T=n_\alpha V^\alpha=0 \label{static condition}
\end{equation}
Using this and that, the total time derivative of the integral has two parts. The first part reflects that $F$ changes as time evolves, and the second part reflects that the surface changes as time evolves. Therefore 
\begin{align}
\frac{d}{dt}&\int_\mathbb{S}Fd\mathbb{S}=\int_\mathbb{S}\partial_tFd\mathbb{S}+\int_\mathbb{S}F\partial_td\mathbb{S} \nonumber \\
&=\int_\mathbb{S}(\partial_tF+F\nabla_\alpha V^\alpha-F\nabla_\alpha V^\alpha)d\mathbb{S} \nonumber \\
&+\int_{x_1}...\int_{x_n}F\partial_t\sqrt{g}dx_1...dx_n\nonumber \\
&=\int_\mathbb{S}(\partial_tF+F\nabla_\alpha V^\alpha-F\nabla_\alpha V^\alpha)d\mathbb{S}\nonumber \\
&+\int_{x_1}...\int_{x_n}F(\nabla_\alpha V^\alpha-CB_\alpha^\alpha)\sqrt{g}dx_1...dx_n\nonumber \\
&=\int_\mathbb{S}(\partial_tF+F\nabla_\alpha V^\alpha-F\nabla_\alpha V^\alpha)d\mathbb{S}\nonumber \\
&+\int_\mathbb{S}F(\nabla_\alpha V^\alpha-CB_\alpha^\alpha)d\mathbb{S}\nonumber \\
&=\int_\mathbb{S}(\partial_tF+\nabla_\alpha(FV^\alpha)-V^\alpha\nabla_\alpha F-CFB_\alpha^\alpha)d\mathbb{S} \label{integration theorem derivation}
\end{align} 
By Gauss theorem, the integral from $\nabla_\alpha(FV^\alpha)$ is converted to the contour integral and, according to the (\ref{static condition}) boundary condition, vanishes. If one relaxes the condition that the surface is closed and is bounded by static contour, then according to (\ref{integration theorem derivation}), the integration theorem (\ref{surface integral}) updates as
\begin{equation}
\partial_t\int_\mathbb{S}Fd\mathbb{S}=\int_\mathbb{S} \dot\nabla F d\mathbb{S}-\int_\mathbb{S}  FCB_\alpha^\alpha d\mathbb{S}+\int_\gamma vFd\gamma \label{contour}
\end{equation}
Since we discuss only closed surfaces here, the last term from (\ref{contour}) becomes irrelevant. 

Note that, for calculation of $\partial_t\sqrt{g}$ in (\ref{integration theorem derivation}), we used Jacobi's theorem about taking derivative for determinants ${\partial g/\partial g^{\alpha\beta}=gg_{\alpha\beta}}$ and (\ref{metric derivative}), so 
\begin{align}
\partial_tg=\frac{\partial g}{\partial g^{\alpha\beta}}\partial_tg^{\alpha\beta}&=gg_{\alpha\beta}(\nabla^\alpha V^\beta+\nabla^\beta V^\alpha-2CB^{\alpha\beta})\nonumber \\
&=2g(\nabla_\alpha V^\alpha-CB_\alpha^\alpha) \label{determinant derivative}
\end{align}
From (\ref{determinant derivative}) according to the complex function derivative rule, the identity 
\begin{equation}
\partial_t\sqrt{g}=\frac{1}{2\sqrt{g}}\partial_tg=\sqrt{g}(\nabla_\alpha V^\alpha-CB_\alpha^\alpha)
\end{equation}
trivially follows. Next, using the (\ref{covariant t}) definition written for scalar functional
 \begin{equation}
 \dot\nabla F=\partial_t F-V^\alpha\nabla_\alpha F \label{scalar function}
 \end{equation}
Applying (\ref{scalar function}) to the (\ref{integration theorem derivation}) with taking into account the Gauss theorem and the boundary condition (\ref{static condition}), the integration theorems (\ref{surface integral}, \ref{contour}) trivially follows. This proof we also reported in  \cite{Svintradze2024arxiv}.   

\textit{Shape Dynamics Equations--} Based on the definitions of velocity field and spatiotemporal covariant derivatives, along with integrations theorems, one can derive surface dynamics equations, see \cite{Svintradze2017, Svintradze2018} and for extensions see \cite{Svintradze2020, Svintradze2023}. Instead of repeating the derivation, we will outline the path forward. According to (\ref{surface action}, \ref{space action}), the most generic action is
\begin{equation}
S=\int_\mathbb{S}\frac{\rho V^2}{2}d\mathbb{S}-\int_\mathbb{S}\mathscr{L}_\mathbb{S}d\mathbb{S}-\int_\Omega\mathscr{L}_\Omega d\Omega \label{boundary action}
\end{equation}    
In three-dimensional space, energy per volume is pressure and per area is the tension therefore to reduce the number of letters, we call $\mathscr{L_\mathbb{S}}\equiv\sigma$ and $\mathscr{L}_\Omega\equiv P$ even though densities are in higher than three dimensions and $\sigma, P$ is only reminiscent of the surface tension and the volumetric pressure. Also (\ref{boundary action}) simplifies to (\ref{space action}) considering that $\mathscr{L}_\mathbb{S}\in\mathscr{L}_\Omega$. 

Next, taking into account the minimum action principle, from (\ref{boundary action}) directly follows that
\begin{equation}
\delta\int_\mathbb{S}\frac{\rho V^2}{2}d\mathbb{S}=\delta(\int_\mathbb{S}\sigma d\mathbb{S}+\int_\Omega P d\Omega)
\end{equation}
According to integration theorems (\ref{surface integral}, \ref{space integral}) and calculations of the covariant time derivative (\ref{covariant t}) (for details see \cite{Svintradze2017, Svintradze2018, Svintradze2020, Svintradze2023, Svintradze2024arxiv}) we have  
\begin{align}
\delta_\bot\int_\mathbb{S}\frac{\rho V^2}{2}d\mathbb{S}&=\int_\mathbb{S} \rho C(\dot\nabla C+2V^\alpha\nabla_\alpha C+V^\alpha V^\beta B_{\alpha\beta})d\mathbb{S} \nonumber\\ 
\delta_\Vert\int_\mathbb{S}\frac{\rho V^2}{2}d\mathbb{S}&=\int_\mathbb{S}\rho V_\alpha(\dot\nabla V^\alpha+V^\beta\nabla_\beta V^\alpha \nonumber \\
&-C\nabla^\alpha C-CV^\beta B_\beta^\alpha)d\mathbb{S} \nonumber \\
\delta\int_\mathbb{S}\sigma d\mathbb{S}&=\int_\mathbb{S}(\dot\nabla\sigma-\sigma CB_\alpha^\alpha)d\mathbb{S} \nonumber \\
\delta_\bot\int_\mathbb{S}\sigma d\mathbb{S}&=\int_\mathbb{S}(\partial_t\sigma-\sigma CB_\alpha^\alpha)d\mathbb{S} \nonumber \\
\delta_\Vert\int_\mathbb{S}\sigma d\mathbb{S}&=-\int_\mathbb{S}V_\alpha\nabla^\alpha\sigma d\mathbb{S} \nonumber \\
\delta\int_\Omega P d\Omega&=\delta_\bot\int_\Omega P d\Omega=\int_\Omega \partial_tPd\Omega+\int_\mathbb{S}PCd\mathbb{S} \nonumber
\end{align}
$\delta_\Vert, \delta_\bot$ designate variations in the tangent space and normal direction correspondingly. Considering that normal deformations mainly govern $\partial_t\sigma$ while $V_\alpha\nabla^\alpha\sigma$ is tangential so that $N^i\partial_t\sigma=\partial_tf^i$, then applying Gauss theorem to the normal component and taking into account conservation of mass the surface dynamics equations follow:
\begin{align}
&\dot\nabla\rho+\nabla_\alpha(\rho V^\alpha)=\rho CB_\alpha^\alpha \nonumber \\
&\partial_i[V^i(\rho(\dot\nabla C+2V^\alpha\nabla_\alpha C+V^\alpha V^\beta B_{\alpha\beta})\nonumber \\
&-P+\sigma B_\alpha^\alpha)]=\partial_t P+\partial_i\partial_tf^i \label{MME} \\
&\rho(\dot\nabla V^\alpha+V^\beta\nabla_\beta V^\alpha-C\nabla^\alpha C-CV^\beta B_\beta^\alpha)=-\nabla^\alpha\sigma \nonumber
\end{align}
Here, the first equation describes the conservation of mass, the second equation shows manifold evolution in the normal direction, and the last equations lead to manifold dynamics in tangent space. Since $\alpha$ is the manifold's dimension, the number of the last equation also indicates the dimensions of the tangent space in the case of GR $\alpha=4$. 
Note that in GR, according to Einstein's field equations, we have ten independent equations, while in shape dynamic equations, the number is reduced to six.  

\textit{EH Action According to CMS} -- To demonstrate the effectiveness of CMS in understanding GR, we begin with the kinetic term of the EH action (\ref{EH action}) without any modifications 
\begin{equation}
S=\frac{1}{2k}\int_\mathbb{S}Rd\mathbb{S}, ... \Rightarrow \mathscr{L}_\mathbb{S}=\frac{1}{2k}R \label{Einstein-Hilbert action}
\end{equation}
where $\mathbb{S}$ stands for surface and $g=-\det g_{\alpha\beta}$ because space-time is pseudo-Riemannian.  
Variation of (\ref{Einstein-Hilbert action}) is widely known \cite{carroll2004spacetime}, so we do not repeat it here. Instead, let's take a variation by CMS. For that, we take into account the integration theorem (\ref{surface integral}) then, the variation of (\ref{Einstein-Hilbert action}) is
\begin{equation}
\delta S=\frac{1}{2k}\frac{d}{dt}\int_\mathbb{S}Rd\mathbb{S}=\frac{1}{2k}\int_\mathbb{S}(\dot\nabla R-RCB_\alpha^\alpha)d\mathbb{S} \label{invariant R}
\end{equation}
By expanding (\ref{invariant R}) according to the invariant time derivative of Ricci scalar by the (\ref{scalar function}) equation, we have  
\begin{equation}
\delta S=\frac{1}{2k}\int_\mathbb{S}(\frac{\partial R}{\partial t}-V^\alpha\nabla_\alpha R-RCB_\alpha^\alpha)d\mathbb{S} \label{exact variation}
\end{equation}
Of course, instead of the (\ref{exact variation}) path, we could take a directly invariant derivative of the Ricci scalar by the (\ref{Ricci}). Then,
\begin{align}
\dot\nabla R&=2\dot\nabla B_\alpha^\alpha-(B^{\alpha\beta}\dot\nabla B_{\alpha\beta}+B_{\alpha\beta}\dot\nabla B^{\alpha\beta})\nonumber\\
&=2(\nabla_\alpha\nabla^\alpha C+CB_\alpha^\beta B_\beta^\alpha)\nonumber\\
&-(\nabla^\alpha\nabla_\beta C+CB^\alpha_\gamma B^\gamma_\beta)(\nabla_\alpha\nabla^\beta C+CB_\alpha^\gamma B_\gamma^\beta) \label{derivative of R}
\end{align}
(\ref{derivative of R}) can be significantly simplified, but even without simplifications, (\ref{derivative of R}) shows strong dominance by normal velocities. That is, the variance of the Ricci scalar is strongly dominated by normal velocities, and therefore, it cannot be modeled by intrinsic tensors only. Any other path to the variance is largely model-dependent and, therefore, approximation. 
Next, as an approximation assume the normal velocity greatly dominates over tangent velocity $C>>V^\alpha$, then we have 
\begin{align}
\partial_t R &=\frac{\partial R}{\partial g^{\alpha\beta}} \partial_t  g^{\alpha\beta} \nonumber\\
\frac{\partial R}{\partial g^{\alpha\beta}}&\approx R_{\alpha\beta} \label{palatini}\\
\delta S&\approx\frac{1}{2k}\int_\mathbb{S}(R_{\alpha\beta}\frac{\partial g^{\alpha\beta}}{\partial t}-RCB^{\alpha\beta}g_{\alpha\beta} )d\mathbb{S}
\end{align}
where (\ref{palatini}) follows from Palatini identity with assumed boundary condition. 
Taking into account (\ref{metric derivative}) and proposing that normal velocity largely dominates over tangent ones, then
\begin{equation}
 \partial_tg^{\alpha\beta}\approx 2CB^{\alpha\beta} \label{Hadamard flow}
\end{equation}
Hadamard proposed the equation (\ref{Hadamard flow}) to explain shape dynamics for water drops. We refer to (\ref{Hadamard flow}) as the Hadamard flow. Taking into account (\ref{Hadamard flow}) we obtain 
\begin{equation}
\delta S\approx\frac{1}{2k}\int_\mathbb{S}2CB^{\alpha\beta}(R_{\alpha\beta}-\frac{1}{2}Rg_{\alpha\beta})d\mathbb{S} \label{calculated action}
\end{equation}
Next, let's add a time-invariable matter field that moves the surface. 
 Since the system is restricted by the parametric time-independent $\sigma$, according to the integration theorem (\ref{surface integral}) the variation of the energy term is 
\begin{equation}
\delta\int_\mathbb{S}\sigma d\mathbb{S}=-\int_\mathbb{S}\sigma CB^{\alpha\beta}g_{\alpha\beta} d\mathbb{S} \label{energy}
\end{equation}
Introducing the designation $T_{\alpha\beta}=\sigma g_{\alpha\beta}$ and demanding that (\ref{calculated action}) has to be identical to (\ref{energy}) for every curvature tensor and normal velocity, one obtains Einstein field equation
\begin{equation}
G_{\alpha\beta}=kT_{\alpha\beta} \label{GR}
\end{equation}
However, during the (\ref{Einstein-Hilbert action}--\ref{calculated action}) derivation, we make several approximations. Therefore, the field equation is not exact and has to be corrected. 
Without consecutive approximations (\ref{exact variation}) variation is
\begin{align}
\delta S=\frac{1}{2k}\int_\mathbb{S}(R_{\alpha\beta}(-\nabla^\alpha V^\beta&-\nabla^\beta V^\alpha+2CB^{\alpha\beta}) \nonumber \\
&-V^\alpha\nabla_\alpha R-RCB_\alpha^\alpha)d\mathbb{S} \label{derived action}
\end{align}
and we can still demand that (\ref{derived action}) is identical to (\ref{energy}) energy variation for every curvature tensor and normal velocity, then the generic field equation is 
\begin{align}
&\kappa R_{\alpha\beta}-\frac{1}{2}Rg_{\alpha\beta}-\frac{V^\alpha\nabla_\alpha R}{2CB^{\alpha\beta}}=kT_{\alpha\beta} \label{GGR} \\
&\kappa=1-\frac{\nabla^\alpha V^\beta+\nabla^\beta V^\alpha}{2CB^{\alpha\beta}}=\frac{\partial_tg^{\alpha\beta}}{2CB^{\alpha\beta}} \label{kapa}
\end{align}
The scenario most closely resembling Einstein’s field equations with a cosmological constant is when $\kappa=1$ and 
\begin{equation}
-V^\alpha\nabla_\alpha R/(2CB^{\alpha\beta})=\Lambda g_{\alpha\beta} \label{cosmological constant}
\end{equation}
Then either $C>>V^\alpha$, $V^\alpha=const$ or ${\nabla^\alpha V^\beta+\nabla^\beta V^\alpha=0}\Rightarrow \nabla_\alpha V^\alpha=0$. Assuming that cosmological constant is time invariable than from (\ref{cosmological constant}), follow that   
\begin{align}
B_\alpha^\alpha\approx 0;  C>>V^\alpha \nonumber\\
\Lambda=\lambda\frac{I^\alpha\nabla_\alpha R}{B_\alpha^\alpha}; \nabla_\alpha V^\alpha=0 \label{CMC}
\end{align}
where $\bm{I}$ is some constant tangent unit vector so that $V^\alpha=VI^\alpha$ and $\lambda=V^\cdot/C$. The trivial solution to (\ref{CMC}) is constant mean curvature surfaces
\begin{equation}
B_\alpha^\alpha=\gamma \nabla_\alpha R^\alpha=const
\end{equation} 
where $\gamma=\frac{\lambda}{\Lambda}$ and $R^\alpha=RI^\alpha$. Therefore, possible solutions to field equations with a cosmological constant are constant mean curvature shapes. Note that (\ref{kapa}) can also be approximately solved directly if one takes into account the (\ref{cosmological constant}) condition.  

Note also that without any algebraic manipulations and approximations, taking into account (\ref{invariant R}) and (\ref{energy}), we can also deduce straightforward field equations for EH action that reads
\begin{equation}
\frac{1}{2CB_\gamma^\gamma}\dot\nabla R_{\alpha\beta}- R_{\alpha\beta}=kT_{\alpha\beta} \label{GE}
\end{equation}
(\ref{GE}) is the generalization of Einstein field equations that look completely different from GR equations, though, due to the course of (\ref{palatini}--\ref{calculated action}) approximations simplifies to the (\ref{GR}) field equations. 


\textit{Equations of Surface Motion} --  Because we try to understand EH action from the viewpoint of phenomenology, we neglect space terms in the (\ref{boundary action})  and concentrate only on surface energy density term $\sigma$. We recover space-related terms when needed to understand the global picture explicitly. In previous sections, we largely dealt with the EH action. Here, we assume that the space-time curving is induced by the kinetic energy of the moving manifold that is modeled as the EH action in GR. Without EH modeling, we have  (\ref{boundary action}) that gives (\ref{MME}) surface dynamics equations which, after neglecting ambient space terms, simplify as  
\begin{align}
&\dot\nabla\rho+\nabla_\alpha(\rho V^\alpha)=\rho CB_\alpha^\alpha \label{MMES} \\
&\partial_i[V^i(\rho(\dot\nabla C+2V^\alpha\nabla_\alpha C+V^\alpha V^\beta B_{\alpha\beta})+\sigma B_\alpha^\alpha)]=\partial_i\partial_tf^i  \nonumber\\
&\rho(\dot\nabla V^\alpha+V^\beta\nabla_\beta V^\alpha-C\nabla^\alpha C-CV^\beta B_\beta^\alpha)=-\nabla^\alpha\sigma \nonumber
\end{align} 
Following conservation of energy $\sigma=const$, in EH action $f_i,\sigma=const $ the equations (\ref{MMES})  further simplify as   
\begin{align}
&\dot\nabla \rho +\nabla_\alpha(\rho V^\alpha)=\rho CB_\alpha^\alpha \nonumber \\
&\rho(\dot\nabla C+2V^\alpha\nabla_\alpha C+V^\alpha V^\beta B_{\alpha\beta})=-\sigma B_\alpha^\alpha \label{Grinfeld equation} \\
&\rho(\dot\nabla V^\alpha+V^\beta\nabla_\beta V^\alpha-C\nabla^\alpha C-CV^\beta B_\beta^\alpha)=0 \nonumber
\end{align}
(\ref{Grinfeld equation}) is known as the Grinfeld equation for dynamic fluid films, and it was proposed to describe the shape dynamics of soap bubbles \cite{PhysRevLett.105.137802}. 

Next, let's assume that (\ref{Grinfeld equation}) is the approximate solution of surface dynamics equations (\ref{MME}) for quasi-constant $P, \sigma, f_i\neq const$ fields. Then for homogenous surfaces $\nabla^i\sigma=0$ and dominantly compressible cases $C>>V^i$, when tangential velocities are infinitesimally small $V^i\approx 0$,  the dynamic manifolds equations (\ref{MME}) simplify as 
\begin{align}
\partial_t\rho&=\rho CB_i^i\nonumber \\
-\partial_i(PV^i)&=\partial_t P+\partial_i\partial_tf^i \label{solved} \\
\rho\partial_tC&=-\sigma B_\alpha^\alpha\nonumber
\end{align}   
Note here that for homogeneous surfaces ${\partial_i P, \partial_i f^i=0}$ and the immediate solution to the second equation, therefore to the system (\ref{solved}) is
\begin{align}
\partial_i V^i&=0 \label{sphere} \\
\partial^2_tC&=-\gamma (\Delta C-2KC)\label{wave}
\end{align}
(\ref{wave}) provides the wave equation for near-planar surfaces where Gaussian curvature $K\approx 0$ is infinitesimally small. It is noteworthy that, in this particular case, even though the system is dominantly compressible, it has an approximate solution (\ref{sphere}) characteristic of dominantly incompressible systems. The first equation gives the spherical solution of
\begin{equation} 
\bm{V}\sim \bm{R}/R^3 \label{spherical solution}
\end{equation} 
where $\bm{R}$ is the position vector. Taking (\ref{spherical solution}) into account, one can extract an analytical solution to (\ref{sphere}) having the form of
\begin{align}
\bm{R}=A(\bm{R}_0+\omega_\alpha R_\alpha \bm{S}^\alpha t)+BR_{0\alpha}e^{\omega_\alpha \bm{S}^\alpha t}+\psi_\alpha\bm{S}^\alpha \label{SF}
\end{align}
where $R_{0\alpha}$ is arbitrary coordinates of the initial position vector $\bm{R}_0$, $\omega_\alpha$ is the coupling constant indicating the frequency of oscillation in $\alpha$ directions, and $\bm{S}^\alpha$ is the vector component of $\bm{S}$ unit vector, $A, B$ are some constants that can be defined by initial conditions, ${e^{\omega_\alpha \bm{S}^\alpha t}=(e^{\omega_x \bm{S}^x t},e^{\omega_y \bm{S}^y t},e^{\omega_z \bm{S}^z t})}$ by definition and ${{\boldsymbol{\psi }}}_{\xi }(\xi ,t)$ is the wave function of the form
\begin{align}
{{\boldsymbol{\psi }}}_{\xi }(\xi ,t)=\sum _{m=1}^{\infty }\,\frac{2}{l}({\int }_{0}^{l}\,{\boldsymbol{\psi }}(\xi ,0)\,\sin \,\frac{m\pi \xi }{l}d\xi ) \nonumber \\
\cdot\cos (\frac{{v}_{0}m\pi t}{l})\sin (\frac{{v}_{0}m\pi t}{l}) \label{S3'}
\end{align}
The exact derivation of the solutions (\ref{sphere}--\ref{S3'}) is given in our works \cite{Svintradze2019, Svintradze2024arxiv}, and we do not repeat it here.  

Note that (\ref{sphere}) and (\ref{wave}) are the solutions of (\ref{MME}, \ref{Grinfeld equation}) for two-dimensional manifolds (surfaces) only.
The same results were obtained about soup bubbles and droplets before \cite{PhysRevLett.105.137802, Svintradze2019}. The combination of analytic (\ref{SF}) and numerical solutions of (\ref{solved}) indicates non-linear oscillations among spherical, oblate/prolate, and non-oblate/prolate shapes, a tendency towards increasing the radius and inducing shape instabilities by forming singularities, the growing amplitude in the oscillations of the surface mass density, meaning increasing the mass instabilities in the surface. Mass instabilities ultimately lead to the development of singularities, which may induce divisions \cite{Svintradze2019}. Increased mass density picks can be considered as model black holes for two-dimensional surfaces. The solutions (\ref{sphere}--\ref{SF}) are explicitly correct only for two manifolds, but the same analyses can be readily extended for higher dimensional surfaces of pseudo-Riemannian geometries. 

\textit{Incompresible spacetime --} Another limit that deserves special attention is $C<<V^\alpha$, that is, interface velocity relative to tangent ones is infinitesimally small, then the space-time can be considered as incompressible or quasi-incompressible $\partial_iV^i\approx 0$. Then, the second equation of (\ref{MME}) further simplifies so that
\begin{align}
V^i\partial_i[\rho V^\alpha V^\beta B_{\alpha\beta}+\sigma B_\alpha^\alpha]&=V^i\partial_iP+\partial_i\partial_tf^i \nonumber\\
\rho V^\alpha V^\beta B_{\alpha\beta}&=P+\partial_if^i-\sigma B_\alpha^\alpha \label{GYL}
\end{align}   
We refer to the solution as the generalized Young-Laplace law for incompressible cases. Assuming $f^i,\sigma\in P\in\Omega$ the (\ref{GYL}) can be rewritten more compactly as
\begin{equation}
P=\rho V^\alpha V^\beta B_{\alpha\beta} \label{GYLC}
\end{equation}    
which has a trivial solution $B_{\alpha\beta}=P(\rho V_\alpha V_\beta)^{-1}$ meaning at appropriate pressure conditions any manifold can spontaneously form. More intriguingly (\ref{GYLC}) leads to generalized Gibbs-Thomson equation
\begin{equation}
T_C=T_0(1-\frac{\mu V^\alpha V^\beta B_{\alpha\beta}}{H}) \label{GGT}
\end{equation} 
here $H$ is the fusion enthalpy of the universe, $\mu$ is molar mass, $T_C$ curvature related melting temperature and $T_0$ planar bath melting temperature \cite{Svintradze2023, Svintradze2024arxiv}. (\ref{GGT}) has a trivial solution of the form
\begin{equation}
B_{\alpha\beta}=\frac{\gamma}{\mu}H_{\alpha\beta}\label{GGTS}
\end{equation}
where $\gamma=(1-T_C/T_0)$ and $H_{\alpha\beta}=H(V_\alpha V_\beta)^{-1}$. (\ref{GYLC}) effectively predicts that the pressure distribution defines the structure of the universe, while (\ref{GGT}-\ref{GGTS}) shows the geometry of the universe if temperature distribution is given. Incidentally, (\ref{GYL}--\ref{GGT}) appears to be incompressible Navier-Stokes equations solutions as well \cite{Svintradze2024arxiv}.  

If one assumes that temperature is uniformly distributed throughout space-time so that $\gamma$ is less than one positive constant and wave propagation velocities in all directions are scaled to the speed of light $c$, then (\ref{GGTS}) simplifies as
\begin{equation}
B_{\alpha\beta}=\frac{\gamma}{\mu c^2}H \delta_{\alpha\beta} \label{GGTSs}
\end{equation}
Again, in the case of GR approximation that the speed of light is the same constant in all directions $V_\alpha=c$, then (\ref{GGTSs}) leads to
\begin{equation} 
B_{\alpha\beta}\simeq \delta_{\alpha\beta} \label{space-time}
\end{equation}
(\ref{space-time}) points out that space-time aims to achieve a locally flat geometry, but this is impossible due to the temperature distribution not being homogeneous and space-time not incompressible. 

\textit{Constant Velocities on Incompressible Spheres --}  Following (\ref{GYLC}), we can calculate surface pressure dynamics on the incompressible, two-dimensional unit sphere for the constant spherical velocities. Even though the calculations are done for two surfaces, its generalizations for any dimensions are straightforward. Due to the manifold dynamics, pressure distribution on the surface is given by
\begin{equation}
P=\rho_0(\omega^2+\varphi^2\sin^2(\omega t+\theta_0)) \label{two pressure}
\end{equation}
where $\rho_0$ is the constant surface mass density, $\omega=V_\theta, \varphi=V_\phi$ are some constants similar to oscillation frequency and amplitude. The appendix Two-sphere Calculations gives the derivation of (\ref{two pressure}). (\ref{two pressure}) indicates that the material point on the unit sphere, with constant spherical $V_\theta, V_\phi=const$ velocities, moves like a wave due to the surface pressures wave-like dynamics induced by the surface dynamics.
 
\textit{Conclusion --} We have proposed a generalization of the GR by CMS's extension of differential geometry. The generalization delivered few striking results. Namely, it indicates that predominantly compressible space-time geometry is neither collapsing nor expanding. It might non-harmonically oscillate between inflation and collapse, frequently developing singularities and mass instabilities that may lead to black hole formation or division. In some approximations, the dominantly compressible space-time simplifies to GR equations by having specific solutions of constant mean curvature shapes. In contrast to predominantly compressible space-time geometry, dominantly incompressible geometry gives a plethora of diverse shapes with different patterns, and dilation strives to form locally flat geometry. As a result, material points with constant spherical velocities on the dominantly incompressible two-sphere start moving like a wave.   

\begin{acknowledgements}
This work is supported by Shota Rustaveli Science Foundation of Georgia by grant no. FR-21-2844 and by Euro Commission’s ERC Erasmus+ collaborative linkage grant between the University of Copenhagen and New Vision University. We also benefit from New Vision University’s internal funding. We want to acknowledge the warm hospitality of the Biocomplexity Department of Niels Bohr Institute, University of Copenhagen, where the paper was completed.
\end{acknowledgements} 

\appendix

\section*{Two-sphere calculations}  
Here, we calculate how the material point moves with constant spherical velocities on the dynamic sphere. The calculations are done only on two-dimensional unit spheres, though extensions in higher dimensions are straightforward. Since we deal with unit spheres, it is convenient to start with spherical coordinates given as
\begin{align}
x(\theta,\phi,t)&=\sin\theta\cos\phi \nonumber \\
y(\theta,\phi,t)&=\sin\theta\sin\phi \nonumber \\
z(\theta,\phi,t)&=\cos\theta \nonumber
\end{align}
The shift, metric, and curvature tensors for the sphere are given as matrices
\begin{equation}
\eta_\alpha^i=\begin{pmatrix}
    \cos\theta\cos\phi&-\sin\theta\sin\phi\\
    \cos\theta\sin\phi&\sin\theta\cos\phi\\
    -\sin\theta&0
    \end{pmatrix} \nonumber
\end{equation}
\begin{equation}
g_{\alpha\beta}=B_{\alpha\beta}=\begin{pmatrix}
    1&0\\
    0&\sin^2\theta\\
    \end{pmatrix};
g^{\alpha\beta}=B^{\alpha\beta}\begin{pmatrix}
    1&0\\
    0&\sin^{-2}\theta\\
    \end{pmatrix} 
    \nonumber
\end{equation}
Note here that the metric tensor and the curvature tensors for spheres are generally different but the same for unit spheres. 

By taking the time derivative of coordinates, the ambient velocity components are calculated as
\begin{align}
V_x&=\frac{dx(\theta,\phi, t)}{dt}=\sin\theta(-\sin\phi)\frac{d\phi}{dt}+\cos\theta\cos\phi\frac{d\theta}{dt} \nonumber \\
V_y&=\cos\theta\sin\phi\frac{d\theta}{dt}+\sin\theta\cos\phi\frac{d\phi}{dt} \nonumber \\
V_z&=-\sin\theta\frac{d\theta}{dt} \nonumber
\end{align}
Next, doting velocity vector $\bm{V}=\begin{pmatrix}V_x&V_y&V_z\end{pmatrix}$ on shift tensor $\eta_\alpha^i$ matrix one gets the first and second components of the tangent velocities, that read
\begin{align}
V_1=V_i\eta_1^i&=(\cos\theta\cos\phi\frac{d\theta}{dt}-\sin\theta\sin\phi\frac{d\phi}{dt})\cos\theta\cos\phi \nonumber\\
&+(\cos\theta\sin\phi\frac{d\theta}{dt}+\sin\theta\cos\phi\frac{d\phi}{dt})\cos\theta\sin\phi \nonumber\\
&+(-\sin\theta\frac{d\theta}{dt})(-\sin\theta)=(\cos^2\theta+\sin^2\theta)\frac{d\theta}{dt}\nonumber\\
&=\frac{d\theta}{dt} \nonumber
\end{align}
\begin{align}
V_2=V_i\eta^i_2&=(\cos\theta\cos\phi\frac{d\theta}{dt}-\sin\theta\sin\phi\frac{d\phi}{dt})(-\sin\theta\sin\phi) \nonumber \\
&+(\cos\theta\sin\phi\frac{d\theta}{dt}+\sin\theta\cos\phi\frac{d\phi}{dt})(\sin\theta\cos\phi) \nonumber \\
&=(\sin^2\theta\sin^2\phi+\sin^2\theta\cos^2\phi)\frac{d\phi}{dt} \nonumber \\
&=\sin^2\theta\frac{d\phi}{dt}  \nonumber
\end{align}
doting $V_1,V_2$ on curvature tensor for the sphere $B^{\alpha\beta}$ and taking into account summation convention we get
\begin{align}
V_\alpha V_\beta B^{\alpha\beta}&=V_1^2B^{11}+V_2^2B^{22} \nonumber\\
&=V_\theta^2+V_\phi^2\sin^2\theta \nonumber
\end{align}
where $V_\theta=d\theta/dt, V_\phi=d\phi/dt$ are spherical velocities. Note here that because the surface moves with constant spherical velocities $\theta=\theta_0+V_\theta t$. Therefore, for constant surface mass density $\rho=\rho_0=const$ the surface pressure changes as
\begin{equation}
P=\rho_0(V_\theta^2+V_\phi^2\sin^2(\theta_0+V_\theta t)) \nonumber
\end{equation}
that is exactly the wave motion on the spherical surface.

\bibliography{apssamp}

\providecommand{\noopsort}[1]{}\providecommand{\singleletter}[1]{#1}%
\begin{thebibliography}{33}%
\makeatletter
\providecommand \@ifxundefined [1]{%
 \@ifx{#1\undefined}
}%
\providecommand \@ifnum [1]{%
 \ifnum #1\expandafter \@firstoftwo
 \else \expandafter \@secondoftwo
 \fi
}%
\providecommand \@ifx [1]{%
 \ifx #1\expandafter \@firstoftwo
 \else \expandafter \@secondoftwo
 \fi
}%
\providecommand \natexlab [1]{#1}%
\providecommand \enquote  [1]{``#1''}%
\providecommand \bibnamefont  [1]{#1}%
\providecommand \bibfnamefont [1]{#1}%
\providecommand \citenamefont [1]{#1}%
\providecommand \href@noop [0]{\@secondoftwo}%
\providecommand \href [0]{\begingroup \@sanitize@url \@href}%
\providecommand \@href[1]{\@@startlink{#1}\@@href}%
\providecommand \@@href[1]{\endgroup#1\@@endlink}%
\providecommand \@sanitize@url [0]{\catcode `\\12\catcode `\$12\catcode
  `\&12\catcode `\#12\catcode `\^12\catcode `\_12\catcode `\%12\relax}%
\providecommand \@@startlink[1]{}%
\providecommand \@@endlink[0]{}%
\providecommand \url  [0]{\begingroup\@sanitize@url \@url }%
\providecommand \@url [1]{\endgroup\@href {#1}{\urlprefix }}%
\providecommand \urlprefix  [0]{URL }%
\providecommand \Eprint [0]{\href }%
\providecommand \doibase [0]{https://doi.org/}%
\providecommand \selectlanguage [0]{\@gobble}%
\providecommand \bibinfo  [0]{\@secondoftwo}%
\providecommand \bibfield  [0]{\@secondoftwo}%
\providecommand \translation [1]{[#1]}%
\providecommand \BibitemOpen [0]{}%
\providecommand \bibitemStop [0]{}%
\providecommand \bibitemNoStop [0]{.\EOS\space}%
\providecommand \EOS [0]{\spacefactor3000\relax}%
\providecommand \BibitemShut  [1]{\csname bibitem#1\endcsname}%
\let\auto@bib@innerbib\@empty
\bibitem [{\citenamefont {Will}(2014)}]{Will:2014aa}%
  \BibitemOpen
  \bibfield  {author} {\bibinfo {author} {\bibfnamefont {C.~M.}\ \bibnamefont
  {Will}},\ }\bibfield  {title} {\bibinfo {title} {The confrontation between
  general relativity and experiment},\ }\href
  {https://doi.org/10.12942/lrr-2014-4} {\bibfield  {journal} {\bibinfo
  {journal} {Living Reviews in Relativity}\ }\textbf {\bibinfo {volume} {17}},\
  \bibinfo {pages} {4} (\bibinfo {year} {2014})}\BibitemShut {NoStop}%
\bibitem [{\citenamefont {Nair}\ \emph {et~al.}(2019)\citenamefont {Nair},
  \citenamefont {Perkins}, \citenamefont {Silva},\ and\ \citenamefont
  {Yunes}}]{PhysRevLett.123.191101}%
  \BibitemOpen
  \bibfield  {author} {\bibinfo {author} {\bibfnamefont {R.}~\bibnamefont
  {Nair}}, \bibinfo {author} {\bibfnamefont {S.}~\bibnamefont {Perkins}},
  \bibinfo {author} {\bibfnamefont {H.~O.}\ \bibnamefont {Silva}},\ and\
  \bibinfo {author} {\bibfnamefont {N.}~\bibnamefont {Yunes}},\ }\bibfield
  {title} {\bibinfo {title} {Fundamental physics implications for
  higher-curvature theories from binary black hole signals in the ligo-virgo
  catalog gwtc-1},\ }\href {https://doi.org/10.1103/PhysRevLett.123.191101}
  {\bibfield  {journal} {\bibinfo  {journal} {Phys. Rev. Lett.}\ }\textbf
  {\bibinfo {volume} {123}},\ \bibinfo {pages} {191101} (\bibinfo {year}
  {2019})}\BibitemShut {NoStop}%
\bibitem [{\citenamefont {Yunes}\ and\ \citenamefont
  {Pretorius}(2009)}]{PhysRevD.80.122003}%
  \BibitemOpen
  \bibfield  {author} {\bibinfo {author} {\bibfnamefont {N.}~\bibnamefont
  {Yunes}}\ and\ \bibinfo {author} {\bibfnamefont {F.}~\bibnamefont
  {Pretorius}},\ }\bibfield  {title} {\bibinfo {title} {Fundamental theoretical
  bias in gravitational wave astrophysics and the parametrized post-einsteinian
  framework},\ }\href {https://doi.org/10.1103/PhysRevD.80.122003} {\bibfield
  {journal} {\bibinfo  {journal} {Phys. Rev. D}\ }\textbf {\bibinfo {volume}
  {80}},\ \bibinfo {pages} {122003} (\bibinfo {year} {2009})}\BibitemShut
  {NoStop}%
\bibitem [{\citenamefont {Abbott}\ \emph {et~al.}(2019)\citenamefont {Abbott}
  \emph {et~al.}}]{PhysRevD.100.104036}%
  \BibitemOpen
  \bibfield  {author} {\bibinfo {author} {\bibfnamefont {B.~P.}\ \bibnamefont
  {Abbott}} \emph {et~al.} (\bibinfo {collaboration} {The LIGO Scientific
  Collaboration and the Virgo Collaboration}),\ }\bibfield  {title} {\bibinfo
  {title} {Tests of general relativity with the binary black hole signals from
  the ligo-virgo catalog gwtc-1},\ }\href
  {https://doi.org/10.1103/PhysRevD.100.104036} {\bibfield  {journal} {\bibinfo
   {journal} {Phys. Rev. D}\ }\textbf {\bibinfo {volume} {100}},\ \bibinfo
  {pages} {104036} (\bibinfo {year} {2019})}\BibitemShut {NoStop}%
\bibitem [{\citenamefont {Dvali}\ \emph {et~al.}(2000)\citenamefont {Dvali},
  \citenamefont {Gabadadze},\ and\ \citenamefont {Porrati}}]{DVALI2000208}%
  \BibitemOpen
  \bibfield  {author} {\bibinfo {author} {\bibfnamefont {G.}~\bibnamefont
  {Dvali}}, \bibinfo {author} {\bibfnamefont {G.}~\bibnamefont {Gabadadze}},\
  and\ \bibinfo {author} {\bibfnamefont {M.}~\bibnamefont {Porrati}},\
  }\bibfield  {title} {\bibinfo {title} {4d gravity on a brane in 5d minkowski
  space},\ }\href
  {https://doi.org/https://doi.org/10.1016/S0370-2693(00)00669-9} {\bibfield
  {journal} {\bibinfo  {journal} {Physics Letters B}\ }\textbf {\bibinfo
  {volume} {485}},\ \bibinfo {pages} {208} (\bibinfo {year}
  {2000})}\BibitemShut {NoStop}%
\bibitem [{\citenamefont {Alexander}\ and\ \citenamefont
  {Yunes}(2009)}]{ALEXANDER20091}%
  \BibitemOpen
  \bibfield  {author} {\bibinfo {author} {\bibfnamefont {S.}~\bibnamefont
  {Alexander}}\ and\ \bibinfo {author} {\bibfnamefont {N.}~\bibnamefont
  {Yunes}},\ }\bibfield  {title} {\bibinfo {title} {Chern--simons modified
  general relativity},\ }\href
  {https://doi.org/https://doi.org/10.1016/j.physrep.2009.07.002} {\bibfield
  {journal} {\bibinfo  {journal} {Physics Reports}\ }\textbf {\bibinfo {volume}
  {480}},\ \bibinfo {pages} {1} (\bibinfo {year} {2009})}\BibitemShut {NoStop}%
\bibitem [{\citenamefont {Nguyen}\ \emph {et~al.}(2023)\citenamefont {Nguyen},
  \citenamefont {Huterer},\ and\ \citenamefont {Wen}}]{PhysRevLett.131.111001}%
  \BibitemOpen
  \bibfield  {author} {\bibinfo {author} {\bibfnamefont {N.-M.}\ \bibnamefont
  {Nguyen}}, \bibinfo {author} {\bibfnamefont {D.}~\bibnamefont {Huterer}},\
  and\ \bibinfo {author} {\bibfnamefont {Y.}~\bibnamefont {Wen}},\ }\bibfield
  {title} {\bibinfo {title} {Evidence for suppression of structure growth in
  the concordance cosmological model},\ }\href
  {https://doi.org/10.1103/PhysRevLett.131.111001} {\bibfield  {journal}
  {\bibinfo  {journal} {Phys. Rev. Lett.}\ }\textbf {\bibinfo {volume} {131}},\
  \bibinfo {pages} {111001} (\bibinfo {year} {2023})}\BibitemShut {NoStop}%
\bibitem [{\citenamefont {Grinfeld}(2013)}]{Grinfeld2013}%
  \BibitemOpen
  \bibfield  {author} {\bibinfo {author} {\bibfnamefont {P.}~\bibnamefont
  {Grinfeld}},\ }\bibinfo {title} {The foundations of the calculus of moving
  surfaces},\ in\ \href {https://doi.org/10.1007/978-1-4614-7867-6_15} {\emph
  {\bibinfo {booktitle} {Introduction to Tensor Analysis and the Calculus of
  Moving Surfaces}}}\ (\bibinfo  {publisher} {Springer New York},\ \bibinfo
  {address} {New York, NY},\ \bibinfo {year} {2013})\ pp.\ \bibinfo {pages}
  {249--265}\BibitemShut {NoStop}%
\bibitem [{\citenamefont {Hadamard}(1908)}]{Hadamard}%
  \BibitemOpen
  \bibfield  {author} {\bibinfo {author} {\bibfnamefont {J.}~\bibnamefont
  {Hadamard}},\ }\href@noop {} {\emph {\bibinfo {title} {M{\'e}moire sur le
  probl{\`e}me d'analyse relatif a l'{\'e}quilibre des plaques {\'e}lastiques
  encastr{\`e}es}}}\ (\bibinfo  {publisher} {Imprimerie nationale},\ \bibinfo
  {year} {1908})\BibitemShut {NoStop}%
\bibitem [{\citenamefont {Thomas}(1961)}]{Thomas}%
  \BibitemOpen
  \bibinfo {editor} {\bibfnamefont {T.~Y.}\ \bibnamefont {Thomas}},\ ed.,\
  \href@noop {} {\emph {\bibinfo {title} {Concepts from Tensor Analysis and
  Differential Geometry}}}\ (\bibinfo  {publisher} {Elsevier},\ \bibinfo {year}
  {1961})\BibitemShut {NoStop}%
\bibitem [{\citenamefont {Grinfeld}(1991)}]{MGrinfeld}%
  \BibitemOpen
  \bibfield  {author} {\bibinfo {author} {\bibfnamefont {M.}~\bibnamefont
  {Grinfeld}},\ }\href@noop {} {\emph {\bibinfo {title} {Thermodynamic methods
  in the theory of heterogeneous systems}}}\ (\bibinfo  {publisher} {Longman Sc
  \& Tech},\ \bibinfo {year} {1991})\BibitemShut {NoStop}%
\bibitem [{\citenamefont {Grinfeld}(2010{\natexlab{a}})}]{Grinfeld2010}%
  \BibitemOpen
  \bibfield  {author} {\bibinfo {author} {\bibfnamefont {P.}~\bibnamefont
  {Grinfeld}},\ }\bibfield  {title} {\bibinfo {title} {Hamiltonian dynamic
  equations for fluid films},\ }\href
  {https://doi.org/10.1111/j.1467-9590.2010.00485.x} {\bibfield  {journal}
  {\bibinfo  {journal} {Studies in Applied Mathematics}\ }\textbf {\bibinfo
  {volume} {125}},\ \bibinfo {pages} {223} (\bibinfo {year}
  {2010}{\natexlab{a}})}\BibitemShut {NoStop}%
\bibitem [{\citenamefont
  {Grinfeld}(2010{\natexlab{b}})}]{PhysRevLett.105.137802}%
  \BibitemOpen
  \bibfield  {author} {\bibinfo {author} {\bibfnamefont {P.}~\bibnamefont
  {Grinfeld}},\ }\bibfield  {title} {\bibinfo {title} {Variable thickness model
  for fluid films under large displacement},\ }\href
  {https://doi.org/10.1103/PhysRevLett.105.137802} {\bibfield  {journal}
  {\bibinfo  {journal} {Phys. Rev. Lett.}\ }\textbf {\bibinfo {volume} {105}},\
  \bibinfo {pages} {137802} (\bibinfo {year} {2010}{\natexlab{b}})}\BibitemShut
  {NoStop}%
\bibitem [{\citenamefont {Grinfeld}(2012)}]{Grinfeld2012}%
  \BibitemOpen
  \bibfield  {author} {\bibinfo {author} {\bibfnamefont {P.}~\bibnamefont
  {Grinfeld}},\ }\bibfield  {title} {\bibinfo {title} {Small oscillations of a
  soap bubble},\ }\href
  {https://doi.org/https://doi.org/10.1111/j.1467-9590.2011.00523.x} {\bibfield
   {journal} {\bibinfo  {journal} {Studies in Applied Mathematics}\ }\textbf
  {\bibinfo {volume} {128}},\ \bibinfo {pages} {30} (\bibinfo {year}
  {2012})}\BibitemShut {NoStop}%
\bibitem [{\citenamefont {Svintradze}(2017{\natexlab{a}})}]{Svintradze2017}%
  \BibitemOpen
  \bibfield  {author} {\bibinfo {author} {\bibfnamefont {D.~V.}\ \bibnamefont
  {Svintradze}},\ }\bibfield  {title} {\bibinfo {title} {Moving manifolds in
  electromagnetic fields},\ }\bibfield  {journal} {\bibinfo  {journal}
  {Frontiers in Physics}\ }\textbf {\bibinfo {volume} {5}},\ \href
  {https://doi.org/10.3389/fphy.2017.00037} {10.3389/fphy.2017.00037} (\bibinfo
  {year} {2017}{\natexlab{a}})\BibitemShut {NoStop}%
\bibitem [{\citenamefont {Svintradze}(2018)}]{Svintradze2018}%
  \BibitemOpen
  \bibfield  {author} {\bibinfo {author} {\bibfnamefont {D.~V.}\ \bibnamefont
  {Svintradze}},\ }\bibfield  {title} {\bibinfo {title} {Closed, two
  dimensional surface dynamics},\ }\bibfield  {journal} {\bibinfo  {journal}
  {Frontiers in Physics}\ }\textbf {\bibinfo {volume} {6}},\ \href
  {https://doi.org/10.3389/fphy.2018.00136} {10.3389/fphy.2018.00136} (\bibinfo
  {year} {2018})\BibitemShut {NoStop}%
\bibitem [{\citenamefont {Svintradze}(2019)}]{Svintradze2019}%
  \BibitemOpen
  \bibfield  {author} {\bibinfo {author} {\bibfnamefont {D.~V.}\ \bibnamefont
  {Svintradze}},\ }\bibfield  {title} {\bibinfo {title} {Shape dynamics of
  bouncing droplets},\ }\href {https://doi.org/10.1038/s41598-019-42580-5}
  {\bibfield  {journal} {\bibinfo  {journal} {Scientific Reports}\ }\textbf
  {\bibinfo {volume} {9}},\ \bibinfo {pages} {6105} (\bibinfo {year}
  {2019})}\BibitemShut {NoStop}%
\bibitem [{\citenamefont {Svintradze}(2020{\natexlab{a}})}]{Svintradze2020}%
  \BibitemOpen
  \bibfield  {author} {\bibinfo {author} {\bibfnamefont {D.~V.}\ \bibnamefont
  {Svintradze}},\ }\bibfield  {title} {\bibinfo {title} {Generalization of the
  kelvin equation for arbitrarily curved surfaces},\ }\href
  {https://doi.org/https://doi.org/10.1016/j.physleta.2020.126412} {\bibfield
  {journal} {\bibinfo  {journal} {Physics Letters A}\ }\textbf {\bibinfo
  {volume} {384}},\ \bibinfo {pages} {126412} (\bibinfo {year}
  {2020}{\natexlab{a}})}\BibitemShut {NoStop}%
\bibitem [{\citenamefont {Svintradze}(2023{\natexlab{a}})}]{Svintradze2023}%
  \BibitemOpen
  \bibfield  {author} {\bibinfo {author} {\bibfnamefont {D.~V.}\ \bibnamefont
  {Svintradze}},\ }\bibfield  {title} {\bibinfo {title} {Generalization of
  young-laplace, kelvin, and gibbs-thomson equations for arbitrarily curved
  surfaces},\ }\href
  {https://doi.org/https://doi.org/10.1016/j.bpj.2023.01.028} {\bibfield
  {journal} {\bibinfo  {journal} {Biophysical Journal}\ }\textbf {\bibinfo
  {volume} {122}},\ \bibinfo {pages} {892} (\bibinfo {year}
  {2023}{\natexlab{a}})}\BibitemShut {NoStop}%
\bibitem [{\citenamefont {Svintradze}(2024)}]{Svintradze:2024aa}%
  \BibitemOpen
  \bibfield  {author} {\bibinfo {author} {\bibfnamefont {D.~V.}\ \bibnamefont
  {Svintradze}},\ }\bibfield  {title} {\bibinfo {title} {Shape dynamics driving
  force for living matter formation},\ }\href
  {https://doi.org/10.1016/j.bpj.2023.11.1489} {\bibfield  {journal} {\bibinfo
  {journal} {Biophysical Journal}\ }\textbf {\bibinfo {volume} {123}},\
  \bibinfo {pages} {236a} (\bibinfo {year} {2024})}\BibitemShut {NoStop}%
\bibitem [{\citenamefont {Svintradze}(2023{\natexlab{b}})}]{Svintradze2023a}%
  \BibitemOpen
  \bibfield  {author} {\bibinfo {author} {\bibfnamefont {D.~V.}\ \bibnamefont
  {Svintradze}},\ }\bibfield  {title} {\bibinfo {title} {Pattern formation on
  dynamic membranes},\ }\href
  {https://doi.org/https://doi.org/10.1016/j.bpj.2022.11.2010} {\bibfield
  {journal} {\bibinfo  {journal} {Biophysical Journal}\ }\textbf {\bibinfo
  {volume} {122}},\ \bibinfo {pages} {364a} (\bibinfo {year}
  {2023}{\natexlab{b}})}\BibitemShut {NoStop}%
\bibitem [{\citenamefont {Svintradze}(2022)}]{Svintradze2022a}%
  \BibitemOpen
  \bibfield  {author} {\bibinfo {author} {\bibfnamefont {D.~V.}\ \bibnamefont
  {Svintradze}},\ }\bibfield  {title} {\bibinfo {title} {Membranes and
  invisibility cloaks},\ }\href
  {https://doi.org/https://doi.org/10.1016/j.bpj.2021.11.2366} {\bibfield
  {journal} {\bibinfo  {journal} {Biophysical Journal}\ }\textbf {\bibinfo
  {volume} {121}},\ \bibinfo {pages} {70a} (\bibinfo {year}
  {2022})}\BibitemShut {NoStop}%
\bibitem [{\citenamefont {Svintradze}(2021)}]{Svintradze2021a}%
  \BibitemOpen
  \bibfield  {author} {\bibinfo {author} {\bibfnamefont {D.~V.}\ \bibnamefont
  {Svintradze}},\ }\bibfield  {title} {\bibinfo {title} {Generalization of the
  gibbs-thomson equation and predicting melting temperatures of
  biomacromolecules in confined geometries},\ }\href
  {https://doi.org/https://doi.org/10.1016/j.bpj.2020.11.1379} {\bibfield
  {journal} {\bibinfo  {journal} {Biophysical Journal}\ }\textbf {\bibinfo
  {volume} {120}},\ \bibinfo {pages} {201a} (\bibinfo {year}
  {2021})}\BibitemShut {NoStop}%
\bibitem [{\citenamefont {Svintradze}(2020{\natexlab{b}})}]{Svintradze2020a}%
  \BibitemOpen
  \bibfield  {author} {\bibinfo {author} {\bibfnamefont {D.~V.}\ \bibnamefont
  {Svintradze}},\ }\bibfield  {title} {\bibinfo {title} {Generalization of the
  kelvin equation and macromolecular surfaces},\ }\href
  {https://doi.org/https://doi.org/10.1016/j.bpj.2019.11.622} {\bibfield
  {journal} {\bibinfo  {journal} {Biophysical Journal}\ }\textbf {\bibinfo
  {volume} {118}},\ \bibinfo {pages} {83a} (\bibinfo {year}
  {2020}{\natexlab{b}})}\BibitemShut {NoStop}%
\bibitem [{\citenamefont {Svintradze}(2017{\natexlab{b}})}]{Svintradze2017a}%
  \BibitemOpen
  \bibfield  {author} {\bibinfo {author} {\bibfnamefont {D.~V.}\ \bibnamefont
  {Svintradze}},\ }\bibfield  {title} {\bibinfo {title} {Geometric diversity of
  living organisms and viruses},\ }\href
  {https://doi.org/https://doi.org/10.1016/j.bpj.2016.11.1676} {\bibfield
  {journal} {\bibinfo  {journal} {Biophysical Journal}\ }\textbf {\bibinfo
  {volume} {112}},\ \bibinfo {pages} {309a} (\bibinfo {year}
  {2017}{\natexlab{b}})}\BibitemShut {NoStop}%
\bibitem [{\citenamefont {Svintradze}(2016)}]{Svintradze2016a}%
  \BibitemOpen
  \bibfield  {author} {\bibinfo {author} {\bibfnamefont {D.~V.}\ \bibnamefont
  {Svintradze}},\ }\bibfield  {title} {\bibinfo {title} {Cell motility and
  growth factors according to differentially variational surfaces},\ }\href
  {https://doi.org/https://doi.org/10.1016/j.bpj.2015.11.3342} {\bibfield
  {journal} {\bibinfo  {journal} {Biophysical Journal}\ }\textbf {\bibinfo
  {volume} {110}},\ \bibinfo {pages} {623a} (\bibinfo {year}
  {2016})}\BibitemShut {NoStop}%
\bibitem [{\citenamefont {Svintradze}(2015)}]{Svintradze2015a}%
  \BibitemOpen
  \bibfield  {author} {\bibinfo {author} {\bibfnamefont {D.~V.}\ \bibnamefont
  {Svintradze}},\ }\bibfield  {title} {\bibinfo {title} {Moving macromolecular
  surfaces under hydrophobic/hydrophilic stress},\ }\href
  {https://doi.org/https://doi.org/10.1016/j.bpj.2014.11.2807} {\bibfield
  {journal} {\bibinfo  {journal} {Biophysical Journal}\ }\textbf {\bibinfo
  {volume} {108}},\ \bibinfo {pages} {512a} (\bibinfo {year}
  {2015})}\BibitemShut {NoStop}%
\bibitem [{\citenamefont {Svintradze}(2014)}]{Svintradze2014a}%
  \BibitemOpen
  \bibfield  {author} {\bibinfo {author} {\bibfnamefont {D.~V.}\ \bibnamefont
  {Svintradze}},\ }\bibfield  {title} {\bibinfo {title} {Conformational motion
  in gene regulatory proteins},\ }\href
  {https://doi.org/https://doi.org/10.1016/j.bpj.2013.11.3606} {\bibfield
  {journal} {\bibinfo  {journal} {Biophysical Journal}\ }\textbf {\bibinfo
  {volume} {106}},\ \bibinfo {pages} {652a} (\bibinfo {year}
  {2014})}\BibitemShut {NoStop}%
\bibitem [{\citenamefont {Svintradze}(2013)}]{Svintradze2013a}%
  \BibitemOpen
  \bibfield  {author} {\bibinfo {author} {\bibfnamefont {D.~V.}\ \bibnamefont
  {Svintradze}},\ }\bibfield  {title} {\bibinfo {title} {Predictive power of
  conformational motion},\ }\href
  {https://doi.org/https://doi.org/10.1016/j.bpj.2012.11.415} {\bibfield
  {journal} {\bibinfo  {journal} {Biophysical Journal}\ }\textbf {\bibinfo
  {volume} {104}},\ \bibinfo {pages} {68a} (\bibinfo {year}
  {2013})}\BibitemShut {NoStop}%
\bibitem [{\citenamefont {Svintradze}(2011)}]{Svintradze2011a}%
  \BibitemOpen
  \bibfield  {author} {\bibinfo {author} {\bibfnamefont {D.~V.}\ \bibnamefont
  {Svintradze}},\ }\bibfield  {title} {\bibinfo {title} {Topology of gene
  delivery systems},\ }\href
  {https://doi.org/https://doi.org/10.1016/j.bpj.2010.12.481} {\bibfield
  {journal} {\bibinfo  {journal} {Biophysical Journal}\ }\textbf {\bibinfo
  {volume} {100}},\ \bibinfo {pages} {52a} (\bibinfo {year}
  {2011})}\BibitemShut {NoStop}%
\bibitem [{\citenamefont {Svintradze}(2009)}]{Svintradze2009a}%
  \BibitemOpen
  \bibfield  {author} {\bibinfo {author} {\bibfnamefont {D.~V.}\ \bibnamefont
  {Svintradze}},\ }\bibfield  {title} {\bibinfo {title} {Conformational motion
  of biological macromolecules},\ }\href
  {https://doi.org/https://doi.org/10.1016/j.bpj.2008.12.3058} {\bibfield
  {journal} {\bibinfo  {journal} {Biophysical Journal}\ }\textbf {\bibinfo
  {volume} {96}},\ \bibinfo {pages} {584a} (\bibinfo {year}
  {2009})}\BibitemShut {NoStop}%
\bibitem [{\citenamefont {{Svintradze}}(2024)}]{Svintradze2024arxiv}%
  \BibitemOpen
  \bibfield  {author} {\bibinfo {author} {\bibfnamefont {D.~V.}\ \bibnamefont
  {{Svintradze}}},\ }\bibfield  {title} {\bibinfo {title} {{Manifold Solutions
  to Navier-Stokes Equations}},\ }\href@noop {} {\bibfield  {journal} {\bibinfo
   {journal} {arXiv e-prints}\ ,\ \bibinfo {eid} {arXiv:2405.15575}} (\bibinfo
  {year} {2024})},\ \Eprint {https://arxiv.org/abs/2405.15575}
  {arXiv:2405.15575 [math.AP]} \BibitemShut {NoStop}%
\bibitem [{\citenamefont {Carroll}\ \emph {et~al.}(2004)\citenamefont
  {Carroll}, \citenamefont {Carroll},\ and\ \citenamefont
  {Addison-Wesley}}]{carroll2004spacetime}%
  \BibitemOpen
  \bibfield  {author} {\bibinfo {author} {\bibfnamefont {S.}~\bibnamefont
  {Carroll}}, \bibinfo {author} {\bibfnamefont {S.}~\bibnamefont {Carroll}},\
  and\ \bibinfo {author} {\bibnamefont {Addison-Wesley}},\ }\href
  {https://books.google.com/books?id=1SKFQgAACAAJ} {\emph {\bibinfo {title}
  {Spacetime and Geometry: An Introduction to General Relativity}}}\ (\bibinfo
  {publisher} {Addison Wesley},\ \bibinfo {year} {2004})\BibitemShut {NoStop}%
\end{thebibliography}%
\end{document}